\documentclass[aps,prl,12pt,preprint,superscriptaddress,showpacs,floatfix]{revtex4-1}
\usepackage{amsmath,amssymb}
\usepackage{graphicx,float}
\usepackage{bm}
\usepackage{mathrsfs}
\usepackage{rotating}
\usepackage[amssymb]{SIunits}

\usepackage{color}

\allowdisplaybreaks[2]

\begin{document}

\title{An opto-magneto-mechanical quantum interface between distant superconducting qubits}

 \author{Keyu \surname{Xia}}  %
  \email{keyu.xia@mq.edu.au}
 \affiliation{ARC Centre for Engineered Quantum Systems, Department of Physics and Astronomy, Macquarie University, NSW 2109, Australia}

 \author{Michael R. \surname{Vanner}}
 \affiliation{School of Mathematics and Physics, The University of Queensland, Brisbane, Queensland 4072, Australia}

 \author{Jason \surname{Twamley}}
 \affiliation{ARC Centre for Engineered Quantum Systems, Department of Physics and Astronomy, Macquarie University, NSW 2109, Australia}

\begin{abstract}
A quantum internet, where widely separated quantum devices are coherently connected, is a fundamental vision for  local and global quantum information networks and processing. 
Superconducting quantum devices can now perform sophisticated  quantum engineering locally on chip  and 
a detailed method to achieve coherent optical quantum interconnection between distant superconducting devices is a vital, but highly challenging, goal.  We describe a concrete opto-magneto-mechanical  system that can interconvert microwave-to-optical quantum information with high fidelity.   In one such node we utilise the magnetic fields generated by the supercurrent of a flux qubit to coherently modulate a mechanical oscillator that is part of a high-Q optical cavity to achieve high fidelity microwave-to-optical quantum information exchange.
We  analyze the transfer between two spatially distant nodes connected by an optical fibre and
using currently accessible parameters we predict that the fidelity of transfer  could be as high as $\sim 80\%$, even with significant loss.
\end{abstract}
\maketitle

\begin{figure*}[h]
 \centering
 \setlength{\unitlength}{0.05\textwidth}
 \begin{picture}(10,10)
 \put(-3,.4){\includegraphics[width=0.9\linewidth]{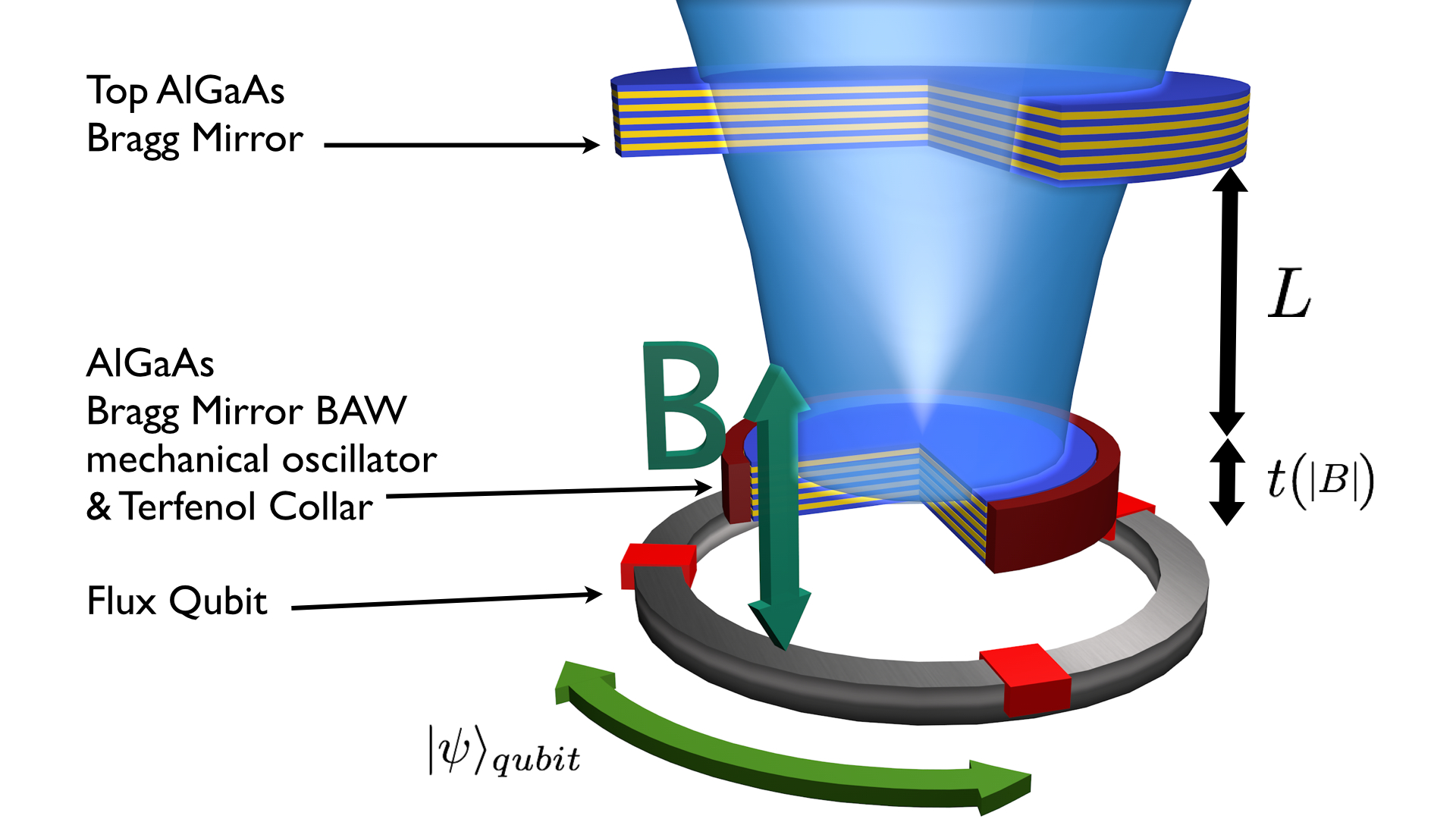}}
 \end{picture}
\caption{\textbf{Schematic for a opto-magneto-mechanical hybrid quantum system for coherent microwave to optical conversion.} Light is injected into a Fabry-Perot micro-cavity that comprises a rigid curved input mirror and a smaller back mirror (cavity length $L$), which is close to a superconducting flux qubit. The back mirror mediates qubit-optical coupling via a bulk-acoustic-wave mechanical resonance in the hundreds of MHz regime. This mechanical mode, where the thickness $t(|B|)$ of the back mirror oscillates, modulates the resonance frequency of the optical cavity and can be resonantly driven by a magnetic field using a surrounding magnetostrictive material such as Terfenol-D.   With our design the quantum state of the superconducting flux qubit can then be made to interact with the optical field.  Here, we consider use of AlGaAs crystalline Bragg mirrors highly reflective at 1,064~nm to simultaneously provide high optical and mechanical quality factors. The whole device of the node is put in a cryogenic environment.}
\label{fig:node}
\end{figure*}

In this report we describe a scheme for a quantum interface between superconducting qubits and optical photons and how to use  such interfaces to coherently transfer quantum information optically between spatially remote superconducting qubits. 
Superconducting quantum devices have recently received a surge of interest and can now perform sophisticated  manipulations on  qubits and microwave photons locally on chip  and are strong candidate for scalable quantum computing \cite{Leo, steffen, Devoret2013}.
A  coherent quantum transfer between distant superconducting qubits can thus serve as the underlying  architecture for a genuine quantum internet. 
The key innovation of our scheme is to employ a magneto-mechanical interaction to provide strong coupling between the quantum state of a superconducting flux qubit (SQ) and a mechanical resonator (MR), which in turn couples via radiation pressure to an optical resonator (OR). 
By driving the SQ we are able to bring it
 into resonance with the magneto-mechanically coupled MR. By also optically driving the optomechanical system we can modulate the optomechanical coupling strength. Using realistic parameters we show that this configuration can achieve a high fidelity coherent swap between the quantum state of the SQ and the OR in an individual node.

 In Fig \ref{fig:node} a detailed design for such an opto-magneto-mechanical system is shown where the optical field in a Fabry Perot cavity is coupled via a Bulk Acoustic Wave mechanical resonance (BAW), which, via a magnetostrictive collar, is coupled magnetically to a SQ.  By coupling two nodes via an optical fibre we create a small quantum network and we demonstrate a protocol to transfer the SQ quantum states  between the  nodes with high  fidelity.

We begin with the quantum interface within a single node for coherent transfer of the superconducting and optically encoded quantum information and following that we analyse coherent quantum transfer between distant superconducting devices. The interface 
 consists of an OR supporting one optical cavity mode and one mechanical vibrational mode and a SQ. The optical cavity mode $\hat{a}$ has a resonance frequency $\omega$ and an intrinsic decay rate $\kappa_i$. The resonance frequency and relaxation rate of mechanical motion mode $\hat{b}$   are $\omega_m$ and $\gamma_m$. To achieve the magneto-mechanical coupling between the SQ and MR we consider that the MR is partly made from a magnetostrictive material like Terfenol-D, which, in the presence of a magnetic field expands in our design, Fig. 1, the magnetostrictive material forms a collar around the BAW mechanical mode to allow magnetic field driving of the mechanical motion. This kind of magnetomechanical coupling has been used for magnetic field sensing and can be quite substantial \cite{Forstner:2012ci}. Indeed this magneto-mechanical coupling is considerably stronger $40-400$~\mega\hertz, than the proposed transfer rates using  piezoelectric  setups $\sim 6$~\mega\hertz  ~\cite{Bochmann:2013di}. Moreover, our design provides excellent mode overlap between the BAW mechanical vibrational mode and the expansion/contraction of the magnetostrictive collar. This should allow even larger magneto mechanical couplings than those previously reported.
The SQ can be modeled as a two-level system (TLS) with the transition frequency $\omega_q$ driven by a classical microwave field with frequency $\omega_d$. The frequency $\omega_q$ can be tuned via a bias magnetic field. 
 The optomechanical coupling can be described by the interaction Hamiltonian $H_{OM}=\hbar g_0 \hat{a}^\dag \hat{a} (\hat{b}^\dag + \hat{b})$, where $\hat{a} (\hat{a}^\dag)$ and $\hat{b} (\hat{b}^\dag)$ are the photon and phonon annihilation (creation) operators for the optical and mechanical modes respectively and $g_{0}$ is the  optomechanical coupling rate. In the frame rotating at the frequency of driving field (on resonance with the SQ), the flux qubit with excited(ground) states $|e\rangle(|g\rangle)$ can be modelled as $H_Q=\hbar\Delta_q /2 \hat{\sigma}_z + \hbar\Omega/2 \hat{\sigma}_x$,  where $\Delta_q=\omega_q-\omega_d$, and the Pauli operators are defined as $\hat{\sigma}_z = |e\rangle \langle e| - |g\rangle \langle g|$ and $\hat{\sigma}_x= |e\rangle \langle g| + |g\rangle \langle e|$. We assume that the flux qubit is driven by a classical field  detuned  by $\Delta_q$ from each SQ's transition frequency and with Rabi frequencies $\Omega$. The macroscopic persistent current $I_p$ circulating within each flux qubit induces a magnetic field $B\propto I_p \hat{\sigma}_z$ dependent on the quantum state of each flux qubit \cite{MCoupling} (see the supplementary information). This magnetic field actuates the motion of the nearby mechanical resonator. This magneto-mechanical interaction is governed by $H_{MQ}=\hbar \Lambda (\hat{b}^\dag + \hat{b}) \hat{\sigma}_z$, where $\Lambda \propto I_p x_{zp}$ is the magnetomechanical-qubit  coupling rate, $x_{zp}=(\hbar/2 m\omega_m)^{1/2}$ is the zero-point width of the mechanical fluctuations with effective mass $m$. 
 At this point we would like to note that while strong coupling between mechanical motion and a superconducting qubit has been achieved in \cite{OConnell}, this has not been achieved with a massive mechanical resonator that also allows coupling to light. The magnetic coupling we are proposing here allows for sufficiently strong qubit-mechanics coupling even for massive mechanical oscillators, such as the dilational mode of Bragg mirror we consider here.

When the flux qubit is put close to a magneto-optomechanical system (coupled MR+OR), they form a three-body quantum system  whose total Hamiltonian is 
\begin{equation}\label{eq:fullH}
\begin{split}
 H_{f}/\hbar = & \Delta'_c \hat{a}^\dag \hat{a} + \omega_m \hat{b}^\dag \hat{b} \\
&+ g_0 \hat{a}^\dag \hat{a}(\hat{b}^\dag + \hat{b}) + \Lambda (\hat{b}^\dag + \hat{b})\hat{\sigma}_z \\
&+ {\textstyle \frac{1}{2}\Omega \hat{\sigma}_x}  
 + (\varepsilon \hat{a} + \varepsilon^* \hat{a}^\dag)\,,
\end{split}
\end{equation}
where the external coherent field $\varepsilon$ drives the optical cavity mode $\hat{a}$ and $\Delta'_c$ is the detuning of this driving from $\hat{a}$.
Setting $\hat{a}\rightarrow\alpha_s +\hat{a}$, and $\hat{b}\rightarrow\beta_s +\hat{b}$, we then linearize the opto-mechanical interaction \cite{Linearization1,Linearization2} and then enter the dressed basis of the flux qubit by $\hat{\sigma}_z \rightarrow \hat{S}_x$, $\hat{\sigma}_x \rightarrow \hat{S}_z$ \cite{dressedbasis} (see supplementary information for the detail of the transformation). The reduced Hamiltonian for the node takes the form 
\begin{equation}\label{eq:efullH}
\begin{split}
 H_{eff}/\hbar= & \Delta_c \hat{a}^\dag \hat{a} + \omega_m \hat{b}^\dag \hat{b} + (G\hat{a}^\dag + G^*\hat{a})(\hat{b}^\dag + \hat{b}) \\
 & + {\textstyle \frac{1}{2}}\Omega \hat{S}_z + \Lambda (\hat{b}^\dag + \hat{b})\hat{S}_x\,,
\end{split}
\end{equation}
with $G=\alpha_s g_0$, and $\alpha_s(\beta_s)$ are the steady-state values of $\langle \hat{a} \rangle(\langle \hat{b} \rangle)$, while  $\Delta_c =\Delta'_c - 2 \eta^2 \omega_m |\alpha_s|^2 $, with the  parameter $\eta = g_0/\omega_m$, $\alpha_s \approx \varepsilon^* /(\Delta_c + i\kappa)$,  $\beta_s \approx - \eta|\alpha_s|^2$ and $\kappa$ is the optical cavity damping rate.  Note that the parameters $\Delta_c, \Omega$ and  $G$ are tunable while $\omega_m$, and $\Lambda$ are not. The total detuning between the flux qubit and the driving is shifted due to the small $\beta_s$ and is given by $\tilde{\Delta}_q=\Delta_q + 2\beta_s\Lambda$. The small cubic term $\hat{a}^\dag \hat{a} (\hat{b}^\dag + \hat{b})$ is negligible because it only causes fast but small oscillation. Throughout our investigation below, we assume that $\tilde{\Delta}_q=0$, but is included in Hamiltonian $H_Q$ for generality.To avoid confusion, here we use different notations for the spin operators of the flux qubit in two different basis. Using this reduced Hamiltonian (\ref{eq:efullH}), we show below that we can reversibly swap a quantum state between the flux qubit and the optical cavity.

\begin{figure}[h]
\centering
\setlength{\unitlength}{0.05\textwidth}
\begin{picture}(10,12)
\put(-6,0){ \includegraphics[width=17cm]{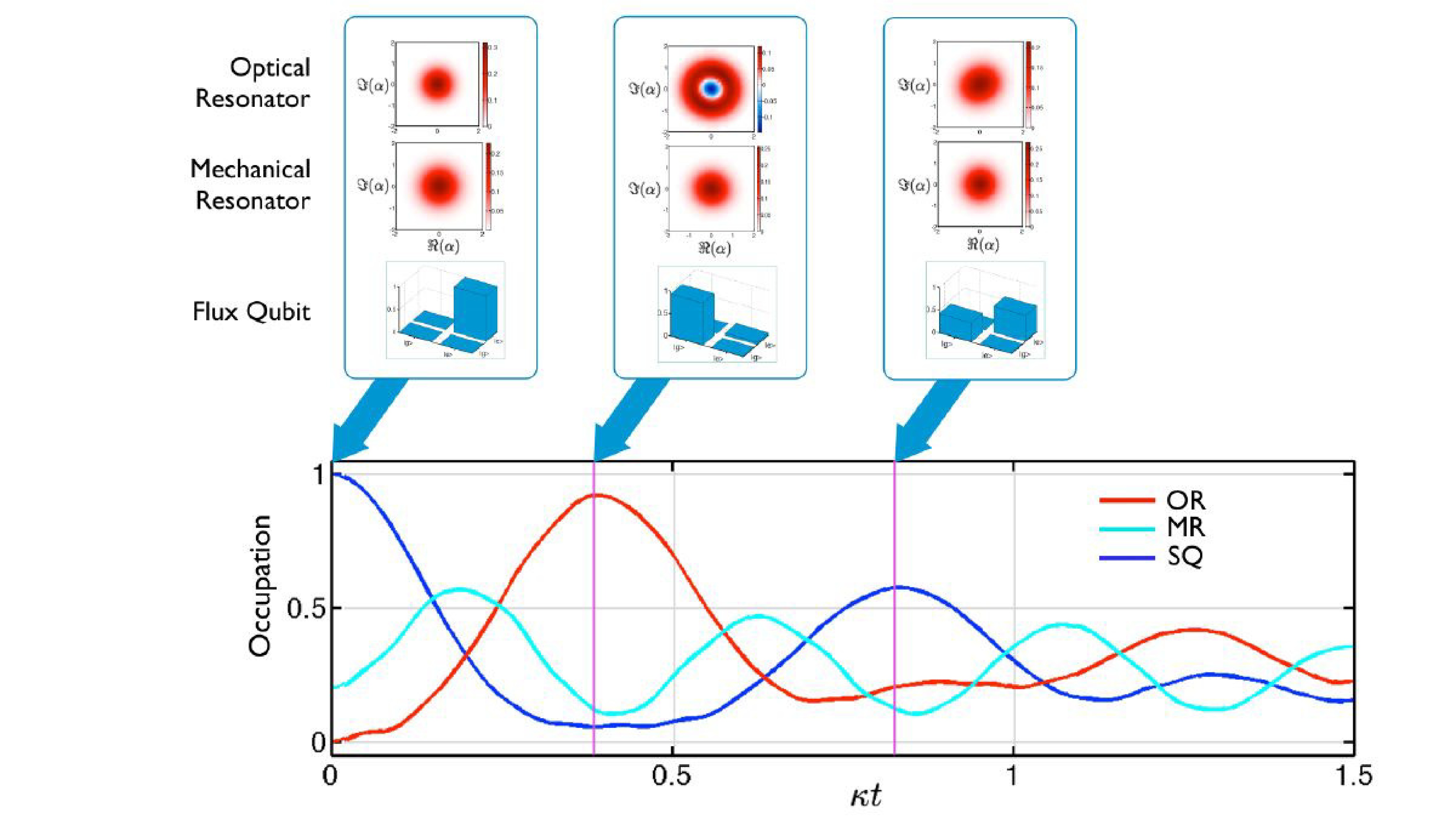}}
\end{picture}
\caption{\textbf{Magneto-mechanically mediated quantum state transfer between a microwave qubit and light.} The main plot shows the Rabi oscillation between an optical resonator (OR) and the flux qubit (SQ)  with small excitation of the mechanical resonator (MR). In addition, we show the states of the various sub-systems at the times $\kappa t=0.0, 0.39, 0.85$, showing the Wigner functions for the OR \& MR and the real part of the reduced density matrix of the SQ (imaginary components are zero).  The system is initially prepared in the state $\rho_i=\rho^{SQ,OR}_i\otimes\rho^{MR}_i$, with $\rho^{SQ,OR}_i$ a pure excited(vacuum) state for the  qubit(optical)  modes, $|\psi_{Q/O}\rangle_i=|e,0\rangle$,  and $\rho^{MR}_i$ is a mechanical thermal state with $\bar{n}=0.2$ phonons. 
{\color{black} We assume no cooling here and the initial phonon occupation is $n_b=0.2$, with $\gamma_q/2\pi=0.02$ \mega\hertz~ and $\kappa/2\pi=10$ \mega\hertz.}
Other parameters are $\Delta_c=\Omega=0.9\omega_m =33\kappa$ and $\Lambda=G=5\kappa$.  The photonic and phonon spaces are truncated to $N_a=4$ and $N_b=6$, respectively. }\label{fig:OneNode}
\end{figure}

We now present our numerical results showing the coherent swapping of the quantum state between the flux qubit and the photon within a single node.
In much of the literature in optomechanics the radiation pressure interaction is linearised as we have done in Eq. (\ref{eq:efullH}), where, via the large optical drive, strong optomechanical coupling can be achieved \cite{Groblacher:2009eh}.
We have numerically verified that this linearisation is accurate for our three-body system, see the supplementary materials. The SQ is initially prepared in the excited state $|\psi_q\rangle_i=|e\rangle$, and the OR  in the photonic vacuum state $|\psi_o\rangle_i=|0\rangle$, and the mechanical resonator is in a thermal state at temperature $T\sim 10$~\milli\kelvin, with $\bar{n}_{th}=0.2$, so that the unwanted decoherence induced by a large mechanical excitation is greatly suppressed. The optical field, mechanical mode and flux qubit interact resonantly and the resulting three-body
 Rabi oscillation is shown in Fig.~\ref{fig:OneNode}. After a half Rabi oscillation, at $\kappa t_1=0.39$, the occupation of the SQ is transferred to the photon with a fidelity of $\mathcal{F}=85\%$ (corresponding to $n_a=0.92$) where $\mathcal{F}=\sqrt{\langle 1| \rho |1\rangle}$. If we continue the evolution the quantum state swaps back to the SQ with $\mathcal{F}=76\%$ at $\kappa t_2=0.85$. {\color{black} If we initially cool the mechanical resonator to $n_b=0$, the fidelity can increase to $\mathcal{F}=89.6\%$ at $\kappa t_2=0.41$ and $\mathcal{F}=78.3\%$ at $\kappa t_2=0.84$.} Our concrete design  uses currently accessible parameters for each of the components and can achieve a high fidelity for microwave-to-optical conversion in a single node. Previous optomechanical 
 \cite{Wang:2012em,Barzanjeh:2012ez}, or solid-state \cite{DiViNCENzo:2011tq}, studies have proposed more abstract schemes for this interconversion, while \cite{Len2}, have demonstrated a classical MW-optical interconversion which operates at $\sim$~ 10\% efficiency.
 
\begin{figure}[h]
\centering
 \setlength{\unitlength}{0.05\textwidth}
 \begin{picture}(10,12)
 \put(-5,.3){\includegraphics[width=0.5\linewidth]{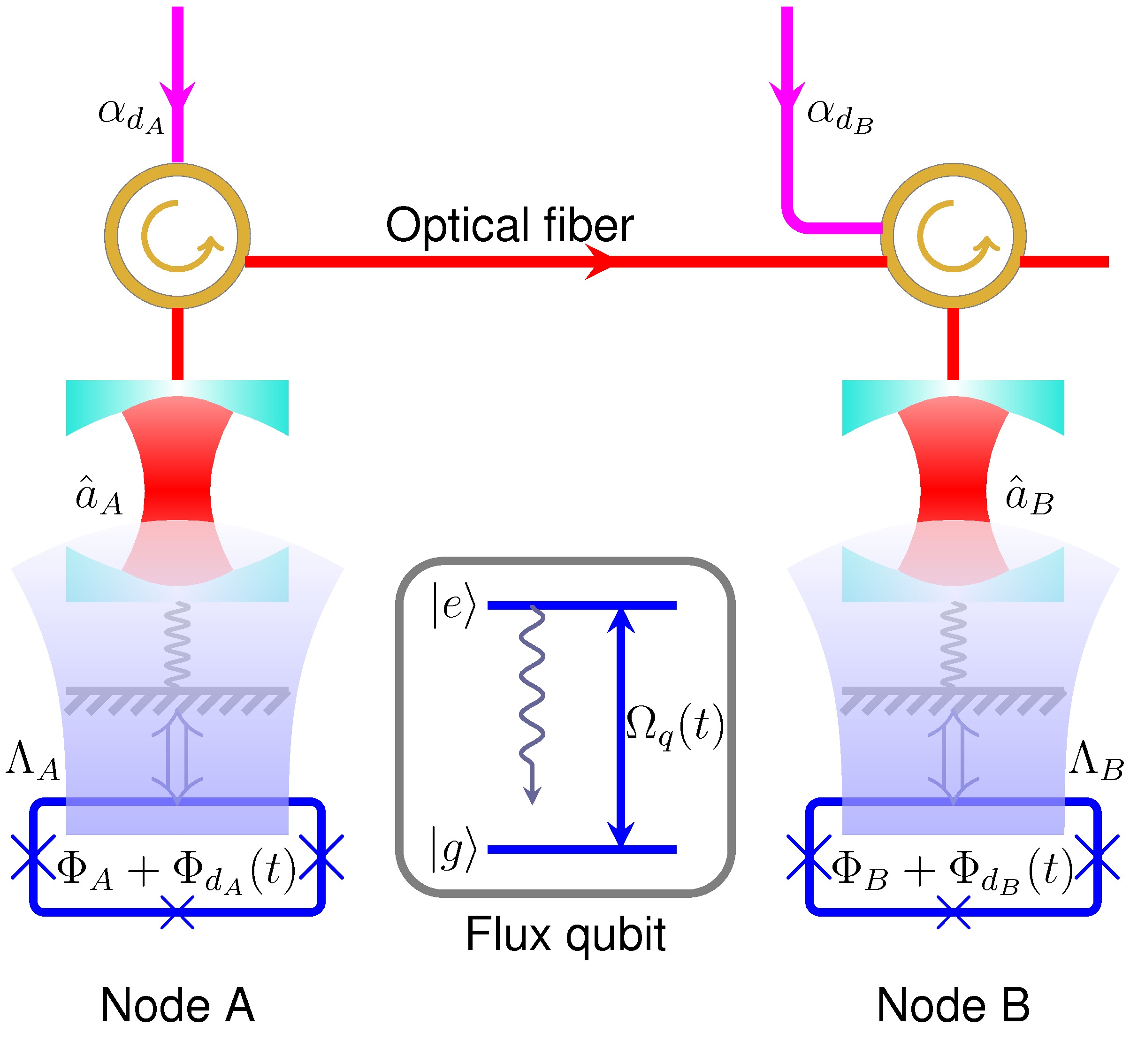}}
  \put(7,.5){\includegraphics[width=0.5\linewidth]{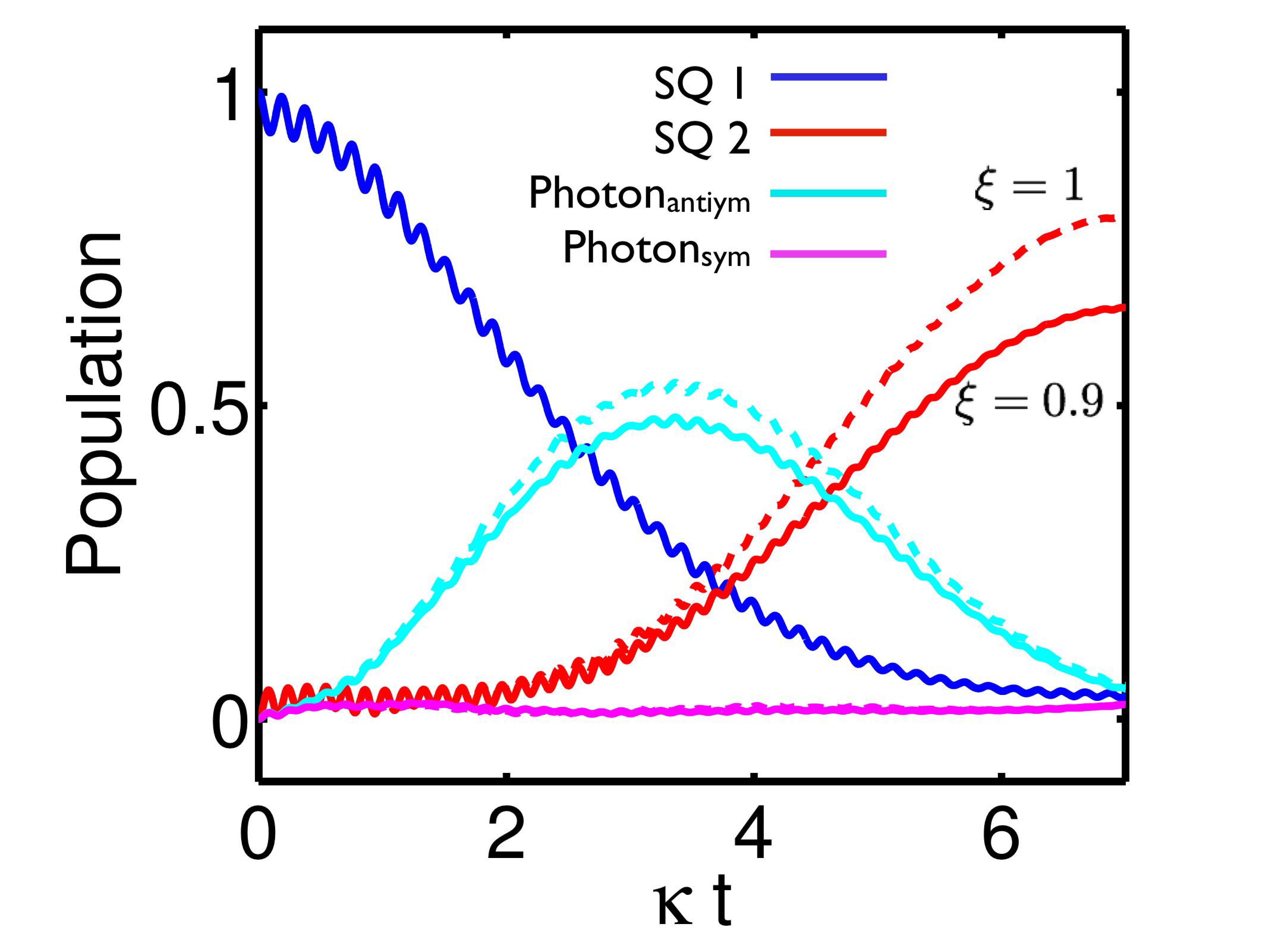}}
  \put(-5.5,9){\bf\Large a}
  \put(+7,8){\bf\Large b}
  \end{picture}
\caption{\textbf{Quantum transfer between two spatially separated superconducting qubits.} (a) Schematic of quantum network between two superconducting flux qubits. Node A and B are identical and each node consists of a flux qubit which is magnetically coupled to a neighbouring optomechanical system. Each flux qubit is coherently driven with a Rabi frequency $\Omega_j,\;j=1,2$, by a time dependent flux bias $\Phi^q_j(t)$, while each optical cavity is coherently driven to coherent states with amplitudes $\alpha_s(\beta_s)$ by input fields (top). By driving the flux qubit we can control the detuning of it's dressed states from the mechanical resonance;
(b) Transport of quantum excitation between distant nodes: The time-dependent population during the evolution of system when $\gamma_q/2\pi=0.02$ \mega\hertz, $\kappa/2\pi = 10$ \mega\hertz, $\omega_m/2\pi=366$ \mega\hertz, mechanical thermal occupation $\bar{n} = 0.2$, and coupling rates $\Lambda=4\kappa, G=1.8\kappa$ and $\Delta_c=\Omega=0.1\omega_m$ requiring an input laser power $P_{in}<16$ \milli\watt.  We assume that the initial phonon number is $\bar{n}\simeq 0.1$ after some pre-cooling.  {\em Dotted Line:}  Lossless transfer between the optical cavities and optical fibre and the latter has no propagation loss, $\xi=1.0$; {\em Solid Line:} Transfer taking account of  photon loss to the environment with  probability $1-\xi=0.1$. The simulation truncated the Hilbert (optical,mechanical) spaces to $(N_a=3, N_b=3)$ which is adequate when transferring a single excitation. 
The blue (red) lines show the population of the excited state $|e\rangle_A$ ($|e\rangle_B$) in node A (B); the cyan and magenta graphs the antisymmetric/symmetric excitation of the cavity modes respectively.
}
\label{fig:setup}
\end{figure}

We now show how to  to perform a coherent  quantum transfer between the flux qubits in two distant nodes connected by an optical fibre. We consider the setup as shown in Fig \ref{fig:setup}(a), where two identical interface nodes  each with separate SQ, MR and OR elements and associated quantities labeled by $j=1,2$.  OR$_j$ couples to a nearby optical fiber with a coupling rate $\kappa_{ex}^{(j)}$ and connects with the other node via this fiber in a cascaded fashion (via the optical circulators). Thus the total decay rate becomes $\kappa_j=\kappa_i^{(j)} + \kappa_{ex}^{(j)}$, and we assume $\kappa_{ex}^{(j)} = \xi_j \kappa_j$ {\color{black}  with $0 \leqslant \xi_j\leqslant 1$. The  $\kappa_{ex}^{(j)}$ denote the couplings into the waveguide connecting the  two distant nodes.} The master equation  describing the cascaded two-node network capturing all forms of dissipation is
 \cite{1997PhRvL..78.3221C,1993PhRvL..70.2269G,1994PhRvA..50.1792G,Stannigel:2010eu,2011PhRvA..84d2341S,2012NJPh...14f3014S,2002PhRvA..65f2308S}
\begin{equation}\label{eq:MENet}
\begin{split}
 \dot{\rho}= & -i(\sum_j H_{eff}^{(j)} \rho -\rho \sum_j H_{eff}^{(j)\dag} )+ \sum_j\mathscr{L}_{noise,j}\rho \\
& + \sum_{i<j}\sqrt{\xi_i\xi_j} \sqrt{\kappa_i \kappa_j}([\hat{a}_j^\dag, \hat{a}_i \rho] + [\rho\hat{a}_i^\dag, \hat{a}_j]) \,,
\end{split}
\end{equation}
where the Lindblad term $\mathscr{L}_{noise,j}\rho=\gamma_A/2(2\hat{A}\rho\hat{A}^\dag - \hat{A}^\dag \hat{A}\rho-\rho \hat{A}^\dag \hat{A})$ with $\gamma_A=\{\gamma_{q_j},\kappa_j\}$ and $\hat{A}=\{\hat{S}_{-,j},\hat{b}_j,\hat{a}_j\}$ describes the open system dynamics of the optomechanical resonators and flux qubits of the $j$th node. Because all interested processes finish within a time much smaller than the ring-down time of the mechanical resonator, we neglect the mechanical decoherence. The second line breaks time-reversal symmetry and models the cascaded quantum transfer from the node $i$ to the node $j$ ($i<j$), or reversely, where $\kappa_j$ is the $j^{th}$ optical cavity decay rate and $\xi_j$ represents photon loss in the transfer (we initially take $\xi_j\approx 1$). 
We can also connect many nodes using the optical fiber as a quantum bus. Such a quantum network allows one to transfer  a quantum state between two distant nodes. We show the simplest case consisting of two nodes only and assume that these two nodes are identical so that $\xi_1=\xi_2$. Currently, a superconducting flux qubit with a decay rate of $\gamma_q/2\pi\sim$\kilo\hertz~ is available using  existing technology \cite{2010PhRvL.105f0503F, 2011NatPh...7..565B, Bishop:2008dc}. 
This decay rate $\gamma_q$ is much smaller than that of the optical cavity. 

The distant transfer of quantum information is shown in Fig.~\ref{fig:setup}(b). We assume a small decay rate for the flux qubit $\gamma_q/2\pi=0.02$ \mega\hertz. When the effective Raman coupling $\zeta=\Lambda G/(\Delta_c-\omega_m)$, is much smaller than the decay rate $\kappa/2$, we find that the constant coupling rates $\Lambda$ and $G$ allow for a transfer of quantum information from  node 1 to  node 2 with a high fidelity. This is an advantage over the complicated time-varying modulation of these rates required in related protocols  \cite{1997PhRvL..78.3221C,1993PhRvL..70.2269G,1994PhRvA..50.1792G,Stannigel:2010eu,2011PhRvA..84d2341S,2012NJPh...14f3014S,2002PhRvA..65f2308S}. Since $\gamma_q \ll \kappa$, it is possible to transfer quantum information between SQs using weak coupling rates $\Lambda$ and $G$. The occupation swapped to the photon from the flux qubit within the donator node quickly transfers to the photon in the acceptor node and then to the acceptor flux qubit. The high-frequency oscillations in the populations are due to the large detuning in the three-body Raman transition. As shown in Fig.~\ref{fig:setup}(b), the state $|e\rangle$ is transferred with the fidelity $\mathcal{F}=0.89$ corresponding to $80\%$ occupation from the qubit $1$ to the qubit $2$ for constant coupling rates $\Lambda=4\kappa, G=1.8\kappa$ and $\Delta_c=0.1\omega_m$ yielding $|\zeta|=0.22\kappa$. The fidelity $\mathcal{F}$ to successfully transfer the quantum state is limited by the decay rate $\gamma_q$ of the flux qubit and is approximately $\mathcal{F}\approx e^{-\gamma_q t_d}$, where $t_d$ is the time to complete the maximal transfer. {\color{black} Therefore, a large $\kappa_{ex}$ but a small $\gamma_q$ is preferable for a high fidelity transfer.} 
We note that during the transfer the symmetric linear combination of both optical  cavity fields  is hardly excited while the antisymmetric one is only weakly excited.
In Fig. ~\ref{fig:setup}(b) we consider the case that all the photonic excitation decaying from the optical cavity A is collected by the optical fiber and then is transferred to the flux qubit in the node B via the optical cavity B as shown in Fig. ~\ref{fig:setup}(b) (dashed).  However, in reality there is a small probability for photon leakage either in the process of cavity-fibre coupling or scattering loss within the fibre itself. In this case, the fidelity of the transfer can be lower. We investigate how well the state can be transferred from  qubit A to  qubit B for $\xi<1$, as shown in Fig. ~\ref{fig:setup}(b) (solid). For a loss corresponding to $\xi=0.9$, we still can transfer the quantum state with $\mathcal{F}\approx 0.81$ corresponding to a probability of $65\%$ from  node A to  node B. In combination with the low temperature, this far detuned Raman transition avoids the unwanted mechanical excitation causing the detrimental decoherence, which destroys the quantum information if using a sequential protocol transferring quantum states from SQ to BAW followed by BAW to optical cavity mode.

All of the requirements to implement our magneto-mechanical superconducting qubit-optical quantum interface can be met with present day technology.
These requirements are primarily: (i) it is desirable to have large mechanical resonance frequency $\omega_m$ to reduce $\bar{n}^{th}_{m}\sim 0.2$ and with pre-cooling to have $\bar{n}_{m}=0.1$ initially. The state transfer fidelity is poor when $\bar{n}_{m}^{th}>1$; (ii) high transfer fidelity requires $G/\kappa\gg 1$ , i.e. strong optomechanical coupling \cite{Groblacher:2009eh,Regal:2011}, and (iii) we must achieve relatively strong magnetic coupling between the mechanics and flux qubit.  Our design is depicted in Fig. \ref{fig:node},  where one end mirror of the FP is constructed of a semiconductor multilayer Bragg circular mirror for light at $\lambda=1064~\nano\meter$,  made from AlGaAs \cite{Cole:2010wx} ($\rho=4200~\kilogram/\meter^3$,  thickness $t=7~\micro\meter$ and radius $r=5~\micro\meter$ \cite{2013PhRvA..87e3818M}. With a Finesse $F\sim 2\times10^5$ and cavity length $L=10~\micro\meter$, the effective mass of the BAW mode $m_{eff}\simeq \pi r^2\,t\rho/3$, the fundamental BAW mode frequency ($\nu_m=v_L/2t$ where $v_L$ is the longitudinal velocity of sound in the semiconductor, $v_L\sim 5130~\meter/\second$,  zero-point motional extent $\Delta x_{zp}^2=\hbar/(2m_{eff}\omega_m)$, temperature $T=10~\milli\kelvin$, and bare optomechanical coupling rate ($g_0=\omega_\lambda\Delta x_{zp}/L$) are: $m_{eff}\sim 770~\pico\gram,\;\omega_m/2\pi\sim 366~\mega\hertz, \;\bar{n}_{m}^{th}\sim 0.2,\;g_0/2\pi\sim 8.3~\kilo\hertz$.  To achieve large enough coupling rates $G/\kappa\sim 1.8$ with a cavity detuning $\Delta_c/\kappa=0.1\omega_m$, we must drive the FP cavity with an optical power $P_{FP}=(G/g_0)^2(\Delta_c^2+\kappa^2)\hbar\omega_\lambda/(2\kappa)\sim 16~\milli{\rm W}$, which is an acceptable power level for such an optical cavity. We note that with these parameters $|\alpha_s|\sim 14,000$. The magneto-mechanical actuation is achieved by using a collar of magneto-strictive material (e.g. Terfenol-D), surrounding the moving mirror. When this collar sits in the local magnetic field of the flux qubit ($|B_{local}|\sim 6.7~ \nano {\rm T})$, the collar exerts stress on the mirror, coupling to the BAW mode. Refs. \cite{Forstner:2012ci, Forstner:2012em}, use a similar arrangement that achieves an extremely large  magneto-mechanical coupling strength $\Gamma/2\pi\sim 1.2~ \giga\hertz$. In our design we assume a more modest coupling strength which is 2-orders of magnitude smaller $\Lambda\sim 4\kappa=2\pi\times 40~\mega\hertz$. This small coupling strength can greatly relax the requirement in the design of the structure. To avoid light interfering with the operation of the flux qubit one may arrange an aperture for the light to pass through contained within the flux qubit loop. Typically, the transition frequency of a flux qubit is much higher than that of a mechanical resonator. This large frequency difference prevents the quantum information exchange between them. To create an effective coupling, a coherent magnetic field $\Phi_{d_j}(t)$ is applied to strongly drive the flux qubit. As a result, the flux qubit oscillates at a Rabi frequency  which matches the mechanical resonance frequency and this enhances their mutual coupling by several orders of magnitude \cite{Forstner:2012ci}. This is physically achievable as a microwave pulse of $P=-112$ dBm propagating in an open transmission line is large enough to drive a flux qubit oscillating at about $\Omega/2\pi=57$~\mega\hertz~ for $I_p=195$~\nano\ampere~\cite{Astafiev:2010cm}. Thus a pulse of $P=-102$ dBm (also used in \cite{Astafiev:2010cm}) can provide a Rabi frequency $570$~\mega\hertz. Therefore, flux qubit Rabi frequencies of hundreds \mega\hertz~ is possible using existing technology. On the other hand, the available enhancement of the zero-point optomechanical coupling is limited by the usable photon number $|\alpha_s|^2$ in the cavity.  Note that the design shown in Fig \ref{fig:node}  is not restricted for operation at a particular optical wavelength and could be suitably modified to operate at the telecommunications wavelength of $1,550~\nano\meter$.

In conclusion, we have proposed and analyzed a magneto-mechanical quantum interface to allow the coherent transfer of quantum information from a superconducting flux qubit to a travelling optical field.
Optical coupling of distant superconducting circuits opens up a new paradigm which may allow for quantum repeaters, distributed quantum computation,  quantum sensing and quantum networking over large spatial scales.

\section*{Acknowledgements}
We would like to thank Garrett D. Cole for useful discussion on crystalline mirror materials. This work was partially funded by the Australian Research Council Centre of Excellence for Engineered Quantum Systems  (EQuS), project number CE110001013 and Discovery Project DP140101638.

\section*{Author Contributions}
K.X.,  M.R.V. and J. T. each contributed to the original concept and the model. K.X. primarily wrote codes for simulations. All three wrote the manuscript.

\section*{Additional Information}
K.X., M.R.V., and J.T. are co-inventors of the magnetic quantum interface described in this Report (internationally protected Australian Patent Application 2014900600).


\bibliographystyle{naturemag}

\end{document}